\documentclass[journal,10pt]{IEEEtran}
\usepackage{tipa}
\usepackage{pifont}\IEEEoverridecommandlockouts
\usepackage{mathrsfs}
\usepackage{amsmath}
\usepackage{amsfonts}
\usepackage{amssymb}
\usepackage{graphicx}
\usepackage{multirow}
\usepackage{cite}
\usepackage{cases}
\usepackage{booktabs}
\usepackage{slashbox}
\usepackage{color}
\usepackage[small]{caption2}
\usepackage[]{caption2}

\usepackage{supertabular}
\usepackage{algorithm}
\usepackage{setspace}
\begin{document}
\begin{spacing}{0.958}
\title {Blocking Probability in Obstructed Tunnels with Reconfigurable Intelligent Surface}

\author{Changshan Chen,~\IEEEmembership{ Member,~IEEE},~Cunhua Pan,~\IEEEmembership{Member, IEEE}
\thanks{C. Chen is with the School of Information Engineering (School of Big Data), Xuzhou University of Technology, Xuzhou 221018, China (e-mail: cschen@xzit.edu.cn).
} 
\thanks{C. Pan is with the School of Electronic Engineering and
Computer Science at Queen Mary University of London, London E1 4NS,
U.K. (e-mail: c.pan@qmul.ac.uk). (\textit{Corresponding author: Cunhua Pan}.) } }
\markboth{}
{}

\maketitle

\begin{abstract}
In the obstructed tunnels, the signal transmission will suffer the risk of ray-path blocking caused by the obstacles owing to the Snell's law. In this letter, the reconfigurable intelligent surface (RIS) that can reflect the electromagnetic waves to any specific directions is introduced to mitigate the signal blocking. The closed-form expressions for blocking probability (BP) for one reflection with single RIS and multiple RISs under various scenarios are derived. Compared with the case without RIS, significant reduction of BP can be found with proper configuration of the RIS. Moreover, the impact of the location of RIS, the height of the transmitter, and the location of the receiver, on the BP is investigated. Finally, the case of multiple obstacles with different distributions is discussed to further verify the effectiveness of RIS on reduction of BP.

\end{abstract}

\begin{IEEEkeywords}
Obstructed tunnel, reconfigurable intelligent surface, blocking probability, multiple RISs.
\end{IEEEkeywords}

\IEEEpeerreviewmaketitle

\section{Introduction}
To meet the demand of excellent wireless data service at anywhere and anytime in emerging wireless networks, the high-reliability and even high-speed data rate wireless communication in tunnels\cite{tunnel_1,tunnel_2,system_2D}, e.g., the subway tunnels, road tunnels and mine tunnels, etc., are required. Nevertheless, limited by the Snell's law,
the obstacles (e.g., the mining equipments and the rockfalls caused by the geological hazards, etc.) in tunnels will result in the ray-path blocking. The mitigation of signal blocking in tunnels with obstacles is crucial to the link reliability, especially for the reliable communication in the challenging environment, such as the rescuing communication for tunnel disaster. To deal with this challenge, the reconfigurable intelligent surface (RIS) is introduced in this letter to reduce the risk of signal blocking in obstructed tunnels.

With configuration of smart controllable reflecting elements, the radio wave can be reflected to any desired direction at the RIS~\cite{cui,zhangrui,RIS_manipulate}. Thus, the RIS can overcome the restriction of Snell's law of reflection to enhance the signal coverage. Utilizing tools from random spatial processes, \cite{reflection_probability} demonstrated that the reflection probability can be significantly improved from below 30\% without RIS to more than 90\% with RIS in free space wireless networks. In recent years, the RIS-assisted communication has attracted substantial attention. Both the prototype measurement \cite{pathloss,dai} and analytical analysis~\cite{zhangrui,RIS_manipulate,RIS_pan} have proved the great potential of RIS in enhancing the wireless communication quality. Most of the contributions on the RIS-aided communication focus on the free space scenarios (like~\cite{RIS_manipulate}, and references therein), while the application of RIS in confined space, like the tunnels, to assist the wireless communication has not been fully explored.
The potential of the application of RIS in underground mine tunnel communication was discussed in \cite{underground_RIS}, nevertheless, the quantitative analysis for this RIS-aided underground communication was not presented. Moreover, it is noted that most of contributions concerning the analysis for RIS-aided communication only pay attention to the single RIS model.
~The outage
probability with multi-RIS was analyzed in \cite{multi_RIS}, and a multi-hop RIS-assisted communication system was employed in \cite{multihop} for the enhancement of network coverage. These works only focus on the multiple RISs-assisted terrestrial communication, the installation of multiple RISs in tunnels to assist wireless communications needs to be further investigated.

In this letter, the closed-form expressions of signal blocking probability (BP) in tunnels with obstacles are derived for both single RIS-aided and multiple RISs-aided (specifically, two RISs) scenario. It is discovered that compared with the case without RIS, the BP can be significantly reduced with proper configuration of RIS. With increased number of RISs, a further decrement of BP is observed.
~Moreover, the impact of the location of RIS, the height of transmitter, and the location of receiver on the BP is discussed to provide more design insights for its practical implementation. Lastly, the investigation for the case of multiple obstacles with different distributions further verifies the effectiveness of RIS on reduction of BP.
\section{System Model}
\begin{figure}[htbp]
 \vspace{-0.5cm}
 \hspace{-0.5cm}
\setlength{\abovecaptionskip}{0pt}
\setlength{\belowcaptionskip}{-15pt}
\center
\includegraphics[width=2.8in,height=1.8in]{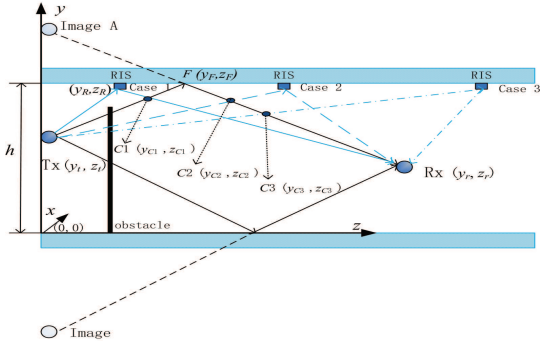}
\caption{System model.} \label{sys}
\end{figure}

We consider a typical straight rectangular tunnel model \cite{system,system_magazine, system_zhou} as shown in Fig. {\ref{sys}}, in which the height of the tunnel is assumed to be $h$ m.
To facilitate the analysis of signal blocking, the two-dimensional model with image ray tracing approach is adopted \cite{system_2D}. Under this model, the considered radio signal transmission is constrained in the $y-z$ plane, and thus the $x$-coordinate is ignored as shown in Fig. {\ref{sys}} (e.g., the coordinate of the origin is expressed as $(0, 0)$ ). Note that this analytical method for signal reflection in $y-z$ plane can be directly applied to the $x-z$ plane.

Assume that only one transmitter Tx and one receiver Rx, the coordinates of which are respectively represented as $(y_t, z_t)$ (${z_t} = 0, 0 < {y_t} < h$)  and $(y_r, z_r)$ ($0 < {y_r} < h, {z_r} > 0$), are configured in the tunnel.
The intersection point of line Rx-A (the image of Tx with respect the ceiling of the tunnel) and the ceiling of the tunnel is $F(y_F, z_F)(y_F=h)$. Then, according to the Snell's law, the radio signal will travel along the ray path Tx-$F$-Rx to the Rx. It should be mentioned that only one reflection is considered here, nevertheless, this image method can also be applicable for analysis of multiple reflections. We consider that a RIS with coordinates $(y_R, z_R)(y_R=h, {z_R} \ge 0)$ is installed on the ceiling of the tunnel to enhance the link reliability. Then, owing to the RIS's ability of manipulating the radio waves \cite{RIS_manipulate}, the radio signal transmission will be no longer restricted by the Snell's law, i.e., the signal can be directly transmitted from the RIS to the Rx \cite{RIS_pan}.

For an example in Fig.~\ref{sys}, if there exists an obstacle in the tunnel, then the signal propagation in one reflection link may be blocked. However, if the RIS is installed on the ceiling of the tunnel (case 1), the radio signal can still be delivered to the Rx along the path Tx-RIS-Rx. Note that for indoor communication, the signal may also be blocked by the building walls, etc., nevertheless, the Tx and Rx are spatially separated in this case. Thus, the signal transmission is different from that of the tunnel communication, in which the Tx and Rx are all installed in the confined space.

Since it is intractable to precisely predict the location and height of the obstacle, then it is supposed that its height, denoted as ${{h_o}}$, is uniformly distributed between $0$ and $h$ and its location along the $z$-axis, denoted as ${{d_o}}$ , follows another uniform distribution in $(0, z_r)$ (note that if the obstacle is located farther than $z_r$, the signal transmission will not be blocked). Therefore, the probability density function (PDF) for ${{h_o}}$ and ${{d_o}}$ can be respectively written as
\begin{equation}
\label{pdf}
{f_{{h_o}}}(y) = \frac{1}{h},{f_{{d_o}}}(z) = \frac{1}{{{z_r}}}.
\end{equation}

To further justify the effectiveness of RIS, the case of multiple obstacles with other distribution models, e.g., the doubly truncated normal distribution (DTND) \cite{truncated} is also considered. The PDF of DTND can be expressed as \cite{truncated} ${f_{DTND}}(x) = \frac{{\varphi (u,{\sigma ^2};x)}}{{\Phi ((b - u)/\sigma ) - \Phi ((a - u)/\sigma )}}$ for $x \in [a,b]$, and ${f_{DTND}}(x)=0$ if $x \notin [a,b]$. $\varphi (u,{\sigma ^2};x) = \frac{{\exp ( - {{(x - u)}^2}/(2{\sigma ^2}))}}{{\sqrt {2\pi } \sigma }}$, and $\Phi (x)$ is the cumulative density function of standard normal distribution.


\vspace{-0.1cm}
\section{Signal Blocking Probability Analysis with Reconfigurable Intelligent Surface}
 In this letter, if the ray path in one reflection is obstructed by the obstacles, then we say this signal is blocked.
The BP analysis for multiple reflections can also be done by employing the above image method~\cite{system_2D}, and we leave it for future work. Note that if one reflection is obstructed, then the line-of-sight path will also be blocked. Therefore, the BP analysis will be helpful for the justification of link reliability in challenging situation of tunnels (e.g., the tunnel collapse, etc.).

As displayed in Fig. {\ref{sys}}, different locations of RIS will result in different BP. Moreover, the BP varies with different numbers of RISs. Thus, in the following, the BP will be calculated for different cases of RIS location and RIS quantity.

Suppose that only one RIS is installed on the ceiling of the tunnel, the BP can be derived in the following theorem.
\newtheorem{theorem}{Theorem}
\begin{theorem}
\label{theo1}
The BP for one reflection in obstructed rectangular tunnel with one RIS (OR) under the system model in (\ref{pdf}) can be expressed as
\begin{align}
{\rm{BP}} = \left\{ \begin{array}{l}
{\rm{Pr[OR|case ~1]}}\\
{\rm{Pr[OR|case~ 2]}}\\
{\rm{Pr[OR|case ~3]}}\\
{\rm{Pr[OR|case~ 4]}}
\end{array} \right.
\end{align}
where ${\rm{Pr[OR|case ~1]}}, {\rm{Pr[OR|case~ 2]}}, {\rm{Pr[OR|case ~3]}}$ and ${\rm{Pr[OR|case~ 4]}}$ are presented at the top of next page, with
\newcounter{TempEqCnt}
\setcounter{TempEqCnt}{\value{equation}}
\setcounter{equation}{2}
 \begin{figure*}[htbp]
 \vspace{-0.8cm}
 \hspace{-1.5cm}
\setlength{\abovecaptionskip}{-5pt}
\setlength{\belowcaptionskip}{-5pt}
\begin{align}
&{\rm{Pr[OR|case~1]}} =C\left[ {(h - {y_t}){z_R} - {k_1}z_R^2 + {{\left( { - {k_2}{z_F} + {k_1}{z_R}} \right)}^2}/\left( {{k_1} - {k_2}} \right)} \right]/2 + C\left[ {\left( {h - {y_r}} \right){z_r} + {z_F}\left( {{y_r} - {y_t}} \right)} \right]/2,\label{pr1}\\
&{\rm{Pr[OR|case~2]}}\!= \!C\!\!\left[ \!{\left( { \!-\! h \!+\! {y_r} \!-\! {k_3}{z_r}}\! \right){z_F} \!+\! \left(\! {{k_2} \!+\! {k_3}}\! \right)z_F^2/2\! +\! {{\left(\! {{y_t} \!-\! {y_r} \!+ \! {k_3}{z_r}}\! \right)}^2}/\left(\! {2\left( {{k_3} \!-\! {k_0}}\! \right)} \!\right)} \!\right] \! \!+\!\!\frac{C}{2}\!\left[ {\left( \!{h \!-\! {y_t}} \right){z_R}\! -\! \left( \!{{z_R} \!-\! {z_r}} \!\right)\left( \!{h \!-\! {y_r}}\! \right)} \right],\label{pr2}\\
&{\rm{Pr[OR|case~3] \!= }}C\left[\! {\left(\! { -\! h \!+\! {y_r} \!-\! {k_3}{z_r}} \!\right){z_F} \!+\! \left(\! {{k_2}\! +\! {k_3}} \right)z_F^2/2}\! \right]\! +\! C\left[ {{{\left( {{y_t} \!-\! {y_r}\! +\! {k_3}{z_r}} \!\right)}^2}/\left( {2\left(\! {{k_3} \!-\! {k_0}}\! \right)} \right)\! +\! \left( {h \!-\! {y_t}}\! \right){z_r}\! -\! {k_0}z_r^2/2} \right],\label{pr3}\\
&{\rm{Pr[OR|case ~4] = }}\!\left\{ \begin{array}{l}
\!\!\!\!\!{C\!\left[\! {\left(\! { -\! h \!+\! {y_r} \!-\! {k_3}{z_r}} \!\right){z_F} \!+\! \left(\! {{k_2}\! +\! {k_3}} \right)z_F^2/2}\! \right]\! +\! C\!\! \left[ {{{\left( {{y_t} \!-\! {y_r}\! +\! {k_3}{z_r}} \!\right)}^2}/\! \! \left( {2\left(\! {{k_3} \!-\! {k_0}}\! \right)} \right)\! +\! \left( {h \!-\! {y_t}}\! \right){z_r}\! -\! {k_0}z_r^2/2}\! \right]}\!,{z_R}\! \le\! {z_N}\\
\!\!\!\!\!C\! \left[ {\left( { - h + {y_r} - {k_3}{z_r}} \right){z_F} \!+ \!\left( {{k_3} + {k_2}} \right)z_F^2/2} \right] +\!  C\! \left[ {\left( {h - {y_r}} \right){z_r} \!+ \!{k_3}z_r^2/2} \!\right],\begin{array}{*{20}{c}}
~~~~~~~~~~~~~~~~~~~~~~~~~{{z_R}\! >\! {z_N}} \label{pr4}
\end{array}
\end{array} \right.
\end{align}
\hrulefill
 \vspace{-0.7cm}
\end{figure*}
\setcounter{equation}{\value{TempEqCnt}}
\setcounter{equation}{6}
$
{\rm{case ~1 = \{ }}{y_t} \ge {y_r}{\rm{ }}~ {\text{or }}~ {y_t} < {y_r}, 0 \le {z_R} \le {z_F}{\rm{\} ,}}~\\
{\rm{case ~2 = \{ }}{y_t} \ge {y_r} ~ {\text{or }}~ {y_t} < {y_r}{{, }}{z_F} < {z_R} \le {z_r}{\rm{\} }},~\\
{\rm{case ~3 = \{ }}{y_t} \ge {y_r}{\rm{, }}{z_R} > {z_r}{\rm{\} }},~\\
{\rm{case ~4 = \{ }}{y_t} < {y_r}{\rm{, }}{z_R} > {z_r}{\rm{\} }},\\
$~ and, $
{k^{'}} = \frac{{{y_r} - 2h + {y_t}}}{{{z_r}}},{z_F} = \left( {h - {y_r} + {k^{'}}{z_r}} \right)/{k^{'}},C = 1/h{z_r},\\
{k_0} = \frac{{h - {y_t}}}{{{z_R}}},{k_1} = \frac{{h - {y_r}}}{{{z_R} - {z_r}}},{k_2} = \frac{{h - {y_t}}}{{{z_F}}},{k_3} = \frac{{{y_r} - h}}{{{z_r} - {z_F}}},\\ {z_N} = \left( {h - {y_r} + {k_4}{z_r}} \right)/{k_4}, {k_4} = \frac{{{y_r} - {y_t}}}{{{z_r} {}}}.
$
\end{theorem}
\itshape {Proof:}  \upshape
See Appendix A.
\hfill $\blacksquare$

Theorem \ref{theo1} reveals the effects of RIS on the reduction of BP in one reflection in the tunnels with obstacles. 
Moreover, it is helpful for the investigation of the impact of system parameters, like the location of RIS, etc., on the BP, which is beneficial for the deployment of RIS in practice to effectively mitigate the risk of signal blocking.

When ${y_t} \to h$, which leads to ${z_F} \to 0$, therefore, the BP can be expressed as (\ref{pr2}) in Theorem \ref{theo1} for ${z_R} \le {z_r}$ (i.e., $ {\rm{Pr[OR|case~2]}}$). For this case,
 the $ {\rm{Pr[OR|case~2]}}$ can be simplified as
  $
 \quad \quad\quad\quad {\rm{Pr[OR|case ~2]}}\mathop  \to \limits^{{y_t} \to h}C\left[ {\left( {{z_r} - {z_R}} \right)\left( {h - {y_r}} \right)/2} \right].
 $

Furthermore, by substituting ${y_t} \to h$ into (\ref{pr4}) under ${{z_R} > {z_N}}$, i.e., the case that without RIS, the BP can be written as
$
{\rm{Pr[OR|case ~4]}}\xrightarrow{{{y_t} \to h,{z_R} > {z_N}}}C\left( {h - {y_r}} \right){z_r}/2.
 $

 Obviously, from the above two equations, the BP aided by RIS is smaller than that without RIS, and the rate of reduction for BP is $C{\left( {h - {y_r}} \right)/2}$.
Hence, we get the following corollary.
\newtheorem{corollary}{Corollary}
\begin{corollary}\label{co}
When ${y_t} \to h$ under ${z_R} \le {z_r}$, the BP decreases linearly with ${z_R}$ at the rate of $C{\left( {h - {y_r}} \right)/2}$.
\end{corollary}

Corollary 1 indicates that, it is better to set the RIS closer to the Rx for the reduction of BP under the conditions mentioned in Corollary 1. If ${y_t} \to h$ and ${z_R} \geqslant {z_r}$, then we have ${\rm{Pr[OR|case~3] \to 0}}$ from (\ref{pr3}), which means the BP can be almost nulled by properly installing the RIS.

When the RIS is located closer to the Tx, for instance, ${z_R} = 0$, the following conclusion can be acquired.
\newtheorem{corollary2}{Corollary}
\begin{corollary}
\label{co2}
When ${z_R} = 0$, the location of Rx along the $z$-axis, i.e., $z_r$, will have no impact on the BP.
\end{corollary}
\itshape {Proof:}  \upshape
By substituting  ${z_R}{\text{ = }}0$ into (\ref{pr1}), we can get
${\rm{Pr[OR|case~1]}}\mathop  = \limits^{{z_R}{\text{ = }}0}  {\left( {h - {y_t}} \right)^2}/\left( {2h\left( {2{y_r} - 3h + {y_t}} \right)} \right) + \left( {h - {y_r}} \right)/2h + \left( {{y_r} - {y_t}} \right)\left( {{y_t} - h} \right)/\left( {2h\left( {{y_r} + {y_t} - 2h} \right)} \right)$, in which no $z_r$ is found.
\hfill $\blacksquare$

From Corollary 2, it is discovered that proper configuration of RIS can make the BP robust to the varying locations of Rx, i.e., the motion of Rx. If it is further assumed that $h \to \infty$, then we can obtain that the BP approaches $1/3$. Nevertheless, 
 for the case without RIS, BP is found to be about $1/2$ if $h \to \infty$. This means that about 30\% of BP can be reduced by the installation of RIS.

The above corollaries along with some simple examples confirm the validity of RIS in mitigating the BP.
In the following, to further validate the effectiveness of RIS on reducing the BP, the BP with multiple RISs will be discussed. 
\newtheorem{theorem1}{Theorem}
\begin{theorem}
\label{theo2}
Suppose that two RISs (TR) are installed in the obstructed rectangular tunnel with location $0 \leqslant {z_{{R_1}}} < {z_F},{z_F} < {z_{{R_2}}}  \leqslant  {z_r}$ under the system model in (\ref{pdf}), then the BP for one reflection is given by
\begin{equation}
{\text{BP}} = {P_{1,TR}} + {P_{2,TR}},
\end{equation}
where,
\begin{align}
  &{P_{1,TR}} = C\left[ {( - {y_t} + h){z_{{R_1}}}/2 - k_1^{{z_{{R_1}}}}z_{{R_1}}^2/2} \right] \nonumber\\
   &+\! C\!\!\left[ {{{\left(\! {k_1^{{z_{{R_1}}}}{z_{{R_1}}}\! -\! {k_2}{z_F}} \!\right)}^2}/\left(\! {2\left( \! {k_1^{{z_{{R_1}}}}\! -\! {k_2}} \!\right)}\! \right) \!+\! {k_2}z_F^2/2} \right],  \\
&{P_{2,TR}} = C\!\!  \left[\! {{{\left(\!  {\!  -\!  {y_r} \! \! +\!\!   {k_3}{z_r} \! +\!  {y_t}} \! \right)}^2}\! /\! \left(\!  {2\left( \! {{k_3} \!\!  -\!\!   k_0^{{z_{{R_2}}}}} \! \right)}\!  \right) \! \! -\!\!   \left( \! {h \! -\!  {y_r}\!  +\!  {k_3}{z_r}}\!   \right){z_F} \! } \right] \nonumber\\
&\! +\!  C\!\!\left[ \! {{k_3}z_F^2\!/2}\!+\!{\left(\! { \!- \!{y_t} \!+ \!h} \right)\!\!{z_{{R_2}}}\! -\!\left( \! {k_0^{{z_{{R_2}}}}z_{{R_2}}^2 \!\! +\! \! \left(\!  {{z_{{R_2}}} \!\!\! -\!\! \! {z_r}}\!  \right)\!\left(\!  {h\!  -\!  {y_r}} \! \right)} \!\right)\!\!/2} \!\right]\!\!,
\end{align}
with
$k_1^{{z_{{R_1}}}} = \frac{{h - {y_r}}}{{{z_{{R_1}}} - {z_r}}}$, $k_0^{{z_{{R_2}}}} = \frac{{h - {y_t}}}{{{z_{{R_2}}}}}$, and ${z_{{R_1}}}, {z_{{R_2}}}$ being the coordinates of the first and second RIS along the $z$-axis, respectively.

\end{theorem}
\itshape {Proof:}  \upshape 
To simplify the analysis, the locations of TR are set to be smaller than $z_r$. Therefore, by employing the similar means of proof for the case 1 and case 2 in Theorem 1, the proof can be completed.
\hfill $\blacksquare$

It should be noted that the BP in Theorem \ref{theo2} is obtained without considering the cooperation between the RISs.
 ~More possible ray-paths will be created with the increased number of RISs, therefore, the BP can be further reduced with two RISs. Further discussion about the benefits of TR will be presented in the numerical results.

Only one obstacle is investigated in above theorems, to be more realistic, let's consider that the number of obstacles, $N>0$ increases linearly with $z_r$ with ratio $k_r>0$, i.e., $N =\left\lceil {{z_r}{k_r}} \right\rceil$ ($\left\lceil  x  \right\rceil$ indicates the smallest integer that is no smaller than $x$). For this case, the BP can be calculated as $\text{BP} = 1 - {(1 - \text{BP})^N}$ with assumption that the obstacles are independent and identically distributed (i.i.d.). If the obstacles follow other distributions, e.g., the DTND, we have the following proposition.
\newtheorem{proposition}{Proposition}
\begin{proposition}
\label{p1}
Considering that the heights of two obstacles with location $0 < {d_{o1}} < {z_R}, {z_R} < {d_{o2}} < {z_{C1}}$ under the case 1 of Theorem 1 are i.i.d. by the DTND in the range $(0, h)$, then the BP can be expressed by
\begin {equation}
{\text{BP}} = 1 - {P_{o1}}{P_{o2}},
\end{equation}
 where,
 \begin{align}
 &{P_{o1}} = {f_{u,\sigma }}({k_0}{d_{o1}} + {y_t})/\left( {{f_{u,\sigma }}(h)} \right),\\
 &{P_{o2}} = {f_{u,\sigma }}({k_1}{d_{o2}} + {y_R} - {k_1}{z_R})/\left( {{f_{u,\sigma }}(h)} \right),
 \end{align}with ${f_{u,\sigma }}(x) = \left( {erf(u/(\sqrt 2 \sigma )) - erf((u - x)/(\sqrt 2 \sigma ))} \right)/2$ and $erf(x) = \frac{2}{{\sqrt \pi  }}\int_0^x {\exp ( - {t^2})} dt$ being the error function.
\end{proposition}

The proof is similar with Theorem 1, where the PDF of obstacles is now changed to ${f_{DTND}}(x)$. Proposition 1 is useful to verify the effectiveness of RIS for non-uniformly distributed multiple obstacles. 

\vspace{0cm}
\section{Numerical Results}
\begin{figure}[t!]
\vspace{-1.5cm}
 \hspace{-1.5cm}
\setlength{\abovecaptionskip}{0pt}
\setlength{\belowcaptionskip}{0pt}
\center
\includegraphics[width=3.5in,height=1.82in]{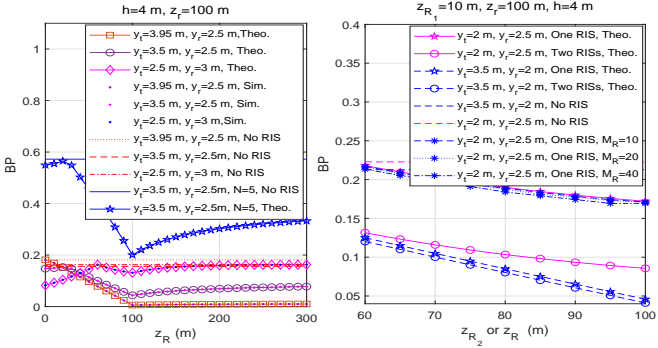}
\caption{The BP versus $z_R$ or $z_{R_2}$ with one RIS (left side) or two RISs (right side).}\label{zR}
\end{figure}

Fig.~\ref{zR} is depicted to investigate the impact of location of RIS on the BP.
 Unless otherwise stated, the $h$ is set to 4 m \cite{height}.   As shown in the left side of Fig.~\ref{zR},
  generally, the BP under one RIS first increases, then decreases, and finally increases with $z_R$. However, when $y_t \to h$, the BP linearly reduces with $z_R$ at a fixed rate shown in the Corollary \ref{co}. Thus, the optimal value of $z_R$ that minimizes the BP varies with the relative height between Tx and Rx. For the case $y_t>y_r$, e.g., $y_t=3.5$ m, $y_r=2.5$ m
   if $z_R$ is set to about 100 m, the BP can be reduced from about 0.16 (without RIS) to the minimum value, about 0.04. If $y_t<y_r$, e.g., $y_t=2.5$ m,  $y_r=3$ m, the BP can be minimized when $z_R$=0, where the BP is reduced to about half of the case without RIS. The above specific examples indicate that proper installation of RIS can effectively mitigate the risk of signal path blocking in tunnels with obstacles. 

  Furthermore, the theoretical results are consistent with the simulations, which verify our analysis. Lastly, if we set $k_r=0.05$, i.e., $N=5$, we can see that, similar with the
   case of one obstacle, by proper installation of RIS, the BP can also be significantly reduced.
    Note that for different cases of $z_R$, the BP is very different according to Theorem 1, therefore, some drastic changes of BP with $z_R$ are found in Fig.~\ref{zR} when $z_R$ changes from one case to another.

From the right side of Fig.~\ref{zR}, if two RISs are installed in the tunnel, the BP can be further decreased. When $z_{R_2}=z_R=100$ m in the scenario $y_t=2$ m, $y_r$=2.5 m, the BP with two RISs is only about 0.08, nevertheless, the BP is doubled with only one RIS. However, it is noted that for some specific cases, e.g., $y_t=3.5$ m, $y_r$=2 m, the BP with two RISs is close to that of the case where only one RIS is configured.

For the RIS with large number of intelligent reflecting meta-surface (IRM), denoted as ${M_R}$, and assume that the channel phases for the channel from Tx to $j$th ($j=1,2,...,M_R$) IRM of RIS $\theta _{tR}^{j}$, and the channel from $j$th IRM of RIS to Rx $\theta _{rR}^{j}$ are known. Then, to reflect the signal to the direction of Rx and to maximize the signal-to-noise ratio (SNR), the adjustable phase induced by $j$th IRM of RIS $\varphi _R^{j}$ can be set as $\varphi _R^{j} = \theta _{tR}^j + \theta _{rR}^{j}$ \cite{zhangrui} under the case that only the cascaded channel Tx-RIS-Rx is found in the tunnels with obstacles.
From right side of Fig.~\ref{zR}, it is detected that the BP is further reduced with the increased ${M_R}$, because more possible paths can be created by the increment of IRM.



Since the signal strength reduces rapidly with the number of reflections in tunnels~\cite{system_zhou}, then the BP can be an indicator for the performance of signal coverage. For instance, suppose that the signal can only be transmitted through the path Tx-RIS-Rx created by the RIS in the obstructed tunnel, then the coverage probability $P(SNR > \gamma )$  can be calculated as $P(SNR > \gamma ) = 1 - BP$ under small value of threshold $\gamma$.

\begin{figure}[t!]
\vspace{-1cm}
 \hspace{-5cm}
\setlength{\abovecaptionskip}{0pt}
\setlength{\belowcaptionskip}{0pt}
\center
\includegraphics[width=3.5in,height=1.82in]{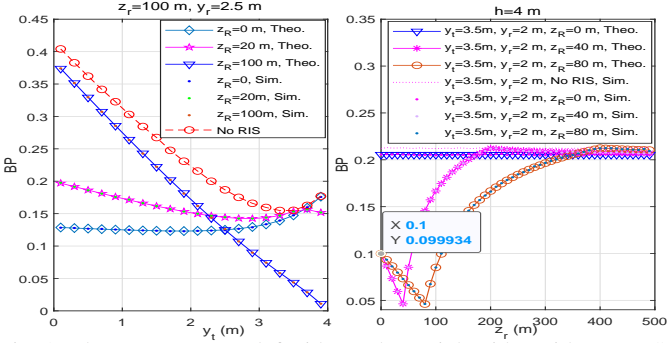}
\caption{The BP versus $y_t$ (left side) and $z_r$ (right side) with one RIS under different cases of $z_R$.}\label{yt}
 \vspace{-0.6cm}
\end{figure}
The impact of height of Tx and the location of Rx (i.e., $z_r$) on the BP is displayed in Fig.~\ref{yt}, which shows that the tendency of BP with $y_t$ is different under different locations of RIS, i.e., $z_R$. If $z_R=0$, the variation of BP with $y_t$ is U-shaped, whereas the BP linearly decreases with $y_t$ if $z_R$=100 m. Moreover, when $z_R=20$ m, the BP first decreases, then increases, and then decreases with $y_t$. This means the optimal value of $y_t$ that minimizes BP varies with $z_R$. Therefore, for different locations of RIS, the proper adjustment for the height of Tx can further reduce BP.

From right side of Fig.~\ref{yt}, it can be seen that for the case without RIS, no impact of $z_r$ on the BP is detected. However, different impacts of $z_r$ on the BP can be found, when only one RIS is installed. If $z_R>0$, the variation of BP with $z_r$ is spoon-shaped. When $z_R=0$, the BP is found to be a constant. Note that, for this special case, there is no significant reduction of BP with RIS. However, from left side of Fig.~\ref{yt}, the height of Rx (i.e., $y_t$) can be adjusted to further reduce BP.  Moreover, it's discovered larger value of $z_R$ can extend the effective range of RIS. For instance, if $z_R=80$ m, the effective range of RIS is (0, 110) under the metric of BP$<0.1$, whereas the effective range of RIS shrinks to about (0, 55) if $z_R=40$ m. The extended effective range of RIS is beneficial for the mobile Rx. Further discussion about installation of RIS with considering the mobility of Rx is left for future work.

\begin{figure}[t!]
\vspace{-1cm}
 \hspace{-1cm}
\setlength{\abovecaptionskip}{0pt}
\setlength{\belowcaptionskip}{0pt}
\center
\includegraphics[width=3in,height=1.82in]{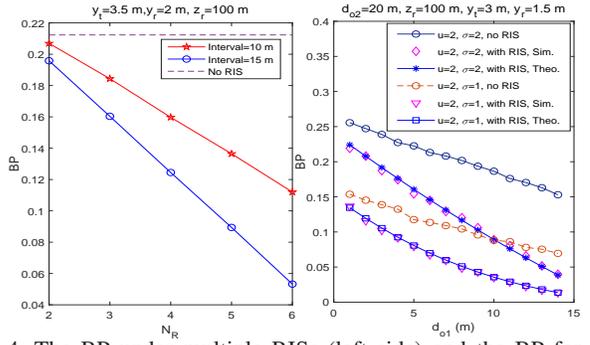}
\caption{The BP under multiple RISs (left side) and the BP for two obstacles with distribution of DTND.}\label{two_obstacles}
 \vspace{-0.6cm}
\end{figure}
The BP with the increased number of RISs (denoted as $N_R$) with different intervals between the RISs is depicted in the left side of Fig.~\ref{two_obstacles}, where a linear decrease of BP with the increasing $N_R$ is discovered.
The BP for two obstacles under distribution of DTND is displayed in right side of Fig.~\ref{two_obstacles} ($h=4$ m, $z_R$= 15 m). We can see that, by employing RIS, the BP can be effectively mitigated, especially for large value of $d_{o1}$ and $\sigma$. This testifies the effectiveness of RIS on the reduction of BP for multiple obstacles with different distributions.


\section{Conclusion}
The BP in tunnels with obstacles by employing RIS is investigated in this letter. Compared with the case without RIS,
 the BP can be significantly reduced with proper configuration of RIS, and the increased number of RISs can lead to a further decrement of BP. Moreover, for different locations of RIS, appropriate adjustment for the height of Tx is beneficial to the mitigation of BP. In addition, a larger distance between RIS and Tx can extend the effective range of RIS under a given BP. Lastly, it's detected the BP can also be effectively reduced for multiple obstacles with different distributions.
 ~ The adjustment of RIS phase shifts for signal enhancement in tunnels is an interesting topic, which will be left for future work.

\appendices
\section{Proof of Theorem 1}
As shown in Fig.~1, if ${y_t} \ge {y_r}$ and ${\rm{ 0}}\le {z_R} \le {z_F}$ (i.e., case 1), then based on the location of the obstacle, four situations, i.e., $0 < {d_o} \le {z_R}$, ${z_R} < {d_o} \le {z_{C1}}$, ${z_{C1}} < {d_o} \le {z_F}$, ${z_F} < {d_o} < {z_r}$, need to be discussed. The point $C1$ is the intersection point of line segment Tx-F and line segment RIS-Rx. When $0 < {d_o} \le {z_R}$, if the obstacle is higher than the intersection point of vertical line $z=d_0$ and line segment Tx-RIS, then the ray path Tx-RIS-Rx will be obstructed. Therefore, for this situation, the BP, denoted as $P_{1,1}$, can be calculated as
\begin{align}
&P_{1,1} ={\Bbb P} (0 < {d_o} \le {z_R},f_{1,1}^{}({d_0}) \le {h_o} \le h) \nonumber \\
 &= \int_0^{{z_R}} {\int_{f_{1,1}^{}(z)}^h {{f_{{h_o}}}(y){f_{{d_o}}}(z)dydz} }\nonumber \\
 & =\! C\!\!\!\int_0^{{z_R}}\!\!\!\!\!\!\!{(\!h \!-\! {y_t}\! +\! {k_0}{z_t} \!- \!{k_0}z)dz}\!=\!C\!\left[\! {(h\! -\! {y_t})\!{z_R} \!- \!{k_0}z_R^2/2} \!\right],
 \vspace{-0.6cm}
\end{align}
where $f_{1,1}^{}(z) = {k_0}z + {y_t}{{ - }}{k_0}{z_t}$ and $C$ is defined in Theorem 1.
Similarly, the BP, for the case ${z_R} < {d_o} \leqslant {z_{C1}}$, denoted as $P_{1,2}$, can be computed as
  \begin{align}
  P_{1,2}& = {\Bbb P}({z_R} < {d_o} \leqslant {z_{C1}},f_{1,2}^{}({d_0}) \leqslant {h_o} \leqslant h) \nonumber \\
  &  = \int_{{z_R}}^{{z_{C1}}} {\int_{f_{1,2}^{}(z)}^h {{f_{{h_o}}}(y){f_{{d_o}}}(z)dydz} }   \nonumber \\
  &  = C\left[ {{k_1}{z_R}\left( {{z_{C1}} - {z_R}} \right) - {k_1}\left( {z_{C1}^2 - z_R^2} \right)/2} \right],
\end{align}
where $f_{1,2}^{}(z) = {k_1}z + {y_R} - {k_1}{z_R}$ and ${z_{C1}} = \frac{{ - {k_2}{z_F} + {k_1}{z_R}}}{{{k_1} - {k_2}}}$.

Following the same method of calculating the $P_{1,1}$ and $P_{1,2}$, the BP $P_{1,3}$ for the case ${z_{C1}} < {d_o} \leqslant {z_F}$ and the BP $P_{1,4}$ for the case ${z_F} < {d_o} < {z_r}$ can be respectively obtained as
\begin{align}
  P_{1,3} & = {\Bbb P}({z_{C1}} < {d_o} \leqslant {z_F},f_{1,3}^{}({d_0}) \leqslant {h_o} \leqslant h)  \nonumber \\
  &  = C\left[ {\left( {{k_2}{z_F}} \right)\left( {{z_F} - {z_{C1}}} \right) - {k_2}\left( {z_F^2 - z_{C1}^2} \right)/2} \right],   \\
   P_{1,4}& = {\Bbb P}({z_F} < {d_o} < {z_r},f_{1,4}^{}({d_0}) \leqslant {h_o} \leqslant h)  \nonumber \\
  &  =\! C\left[\! {\left( {h \!-\! {y_r} \!+\! {k_3}{z_r}}\! \right)\left( {{z_r} \!-\! {z_F}} \right)\! -\! {k_3}\left(\! {z_r^2 \!-\! z_F^2} \!\right)/2} \right],
  \end{align}
 with $f_{1,3}^{}(z) = {k_2}z + {y_F} - {k_2}{z_F}$ and $f_{1,4}^{}(z) = {k_3}z + {y_r} - {k_3}{z_r}$.
 The overall BP for the case 1 is then acquired by combing the above four different cases of BP, i.e.,
 \begin{align}
   {\rm{Pr[OR|case~ 1] }}=P_{1,1}^{} + P_{1,2}^{} + P_{1,3}^{} + P_{1,4}^{}. \label{case11}
 \end{align}
 By substituting $z_t=0$ and $y_R=h$ into (\ref{case11}) and with some manipulations, the equation (\ref{pr1}) can be obtained. Note that for ${y_t} < {y_r}$ and ${\rm{ 0}}\le {z_R} \le {z_F}$, we can get the same expressions.

Similar to case 1, there also exist four situations  in case 2, the BP of which can be calculated as
$
{\rm{Pr[OR|case ~2] = }}P_{2,1}^{} + P_{2,2}^{} + P_{2,3}^{} + P_{2,4}^{},
$
where
\begin{align}
&P_{2,1}\! =\! {\Bbb P}(0 \!<\! {d_o} \leqslant {z_F},f_{2,1}^{}({d_0})\! \leqslant {h_o}\! \leqslant h){\text{\! = \! }}C{k_2}z_F^2/2,\\
  & P_{2,2}^{}{\text{ = }}{\Bbb P}({z_F} < {d_o} \leqslant {z_{C2}},f_{2,2}^{}({d_0}) \leqslant {h_o} \leqslant h)
  \nonumber \\
  &  \! =\!  C\left[ {\left( {h \! -\!  {y_r} \! +\!  {k_3}{z_r}} \right)\left( {{z_{C2}} \! -\!  {z_F}} \right) \! -\!  {k_3}\left( {z_{C2}^2 - z_F^2}\!  \right)/2} \right],\\
  & P_{2,3}^{}{\text{ = }}{\Bbb P}({z_{C2}} < {d_o} \leqslant {z_R},f_{2,3}^{}({d_0}) \leqslant {h_o} \leqslant h)  \nonumber \\
  & {\text{=\! }}C\left[ {\left( {h \! -\!  {y_t}\!  +\!  {k_0}{z_t}} \right)\left( {{z_R} \! -\!  {z_{C2}}} \right)\!  -\!  {k_0}\left( {z_R^2 \! -\!  z_{C2}^2} \right)/2} \right],\\
  & P_{2,4}^{}{\text{ = }}{\Bbb P}({z_R} < {d_o} < {z_r},f_{2,4}^{}({d_0}) \leqslant {h_o} \leqslant h) \nonumber \\
  & {\text{ = }}C\left[ {\left( {h - {y_R} + {k_1}{z_R}} \right)\left( {{z_r} - {z_R}} \right) - {k_1}\left( {z_r^2 - z_R^2} \right)/2} \right],
  \end{align}
  with $f_{2,1}^{}(z) = f_{1,3}(z)$, $f_{2,2}^{}(z) = f_{1,4}(z)$, $f_{2,3}^{}(z) = f_{1,1}(z)$, $f_{2,4}^{}(z) = f_{1,2}(z)$, and ${z_{C2}} = \frac{{{y_t} - {y_r} + {k_3}{z_r} - {k_0}{z_t}}}{{{k_3} - {k_0}}}$. After some manipulations, ${\rm{Pr[OR|case~ 2]}}$ can be simplified as (\ref{pr2}).

  As for case 3 shown in Fig.~\ref{sys}, only three situations need to be investigated, then the BP for this case can be expressed as ${\rm{Pr[OR|case~3] = }}P_{3,1}^{} + P_{3,2}^{} + P_{3,3}^{}$, where
  \begin{align}
  & P_{3,1} =\! {\Bbb P}(0 \!<\! {d_o} \leqslant {z_F},f_{2,1}^{}({d_0}) \!\leqslant {h_o} \!\leqslant h){\text{ \!=\! }}C{k_2}z_F^2/2, \\
  & P_{3,2}^{}{\text{ = }}{\Bbb P}({z_F} < {d_o} \leqslant {z_{C3}},f_{2,2}^{}({d_0}) \leqslant {h_o} \leqslant h) \nonumber \\
  & {\text{ = }}C\left[ {\left(\! {h \!- \!{y_r}\! + \!{k_3}{z_r}} \!\right)\left(\! {{z_{C3}}\! -\! {z_F}} \right) \!-\! {k_3}\left( {z_{C3}^2\! -\! z_F^2}\! \right)/2} \right], \\
  & P_{3,3}^{}{\text{ = }}{\Bbb P}({z_{C3}} < {d_o} < {z_r},f_{3,3}^{}({d_0}) \leqslant {h_o} \leqslant h)  \nonumber \\
  & {\text{ = }}C\left[ {\left(\! {h\! -\! {y_t} \!+ \!{k_0}{z_t}} \right)\left( {{z_r}\! -\! {z_{C3}}} \right) \!-\! {k_0}\left( {z_r^2 \!-\! z_{C3}^2} \right)/2} \right],
  \end{align}
with $f_{3,3}^{}(z) = f_{1,1}^{}(z), {z_{C3}} = {z_{C2}}$. With some manipulations, we can get the simplified form of
${\rm{Pr[OR|case~3]  }}$ in (\ref{pr3}).

\begin{figure}[htbp]
 \vspace{-0.5cm}
 \hspace{-0.5cm}
\setlength{\abovecaptionskip}{0pt}
\setlength{\belowcaptionskip}{-10pt}
\center
\includegraphics[width=3in,height=1.5in]{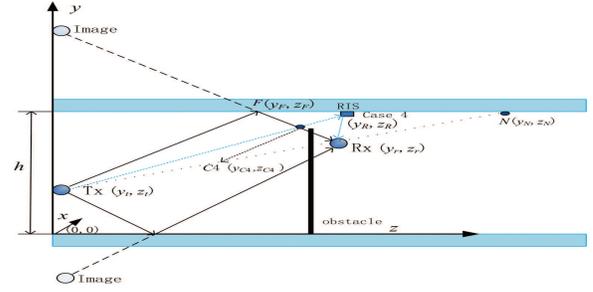}
\caption{The sketch map for case 4.} \label{case4}
\vspace{0cm}
\end{figure}

When regarding the case 4 shown in Fig.~\ref{case4},
the calculations of BP should be divided into two parts, i.e., ${{z_R} \le {z_N}}$ and ${{z_R} > {z_N}}$, where the point $N$ is the intersection point of line Tx-Rx and line $y=h$.
If ${{z_R} \le {z_N}}$, the BP for this case is same as that of case 3, i.e., ${\rm{Pr[OR|case~ 4] = Pr[OR|case ~3]}}$. Nevertheless, if ${{z_R} > {z_N}}$, the installation of RIS will have no contribution on reduction of the BP, since the intersection point of line F-Rx and line Tx-RIS is lower than Rx. The BP of this case is just the same as that of the case without RIS, which can be calculated as
${\rm{Pr[OR|case~4] = }}P_{4,1}^{} + P_{4,2}^{},
$
where
\begin{align}
  & P_{4,1}^{} = {\Bbb P}(0 < {d_o} \leqslant {z_F},f_{4,1}^{}({d_0}) \leqslant {h_o} \leqslant h)  \nonumber \\
  & {\text{ = }}C\left[ {\left( {h - {y_F} + {k_2}{z_F}} \right){z_F} - {k_2}z_F^2/2} \right],  \\
  & P_{4,2}^{} = {\Bbb P}({z_F} < {d_o} < {z_r},f_{4,2}^{}({d_0}) \leqslant {h_o} \leqslant h) \nonumber \\
  & {\text{ = }}C\left[ {\left( {h - {y_r} + {k_3}{z_r}} \right)\left( {{z_r} - {z_F}} \right) - {k_3}\left( {z_r^2 - z_F^2} \right)/2} \right],
  \end{align}
with $f_{4,1}^{}(z)=f_{1,3}^{}(z), f_{4,2}^{}(z) =f_{1,4}^{}(z) $. After some manipulations, the BP of case 4 can be
written as in (\ref{pr4}).

\tiny
\vspace{-0.1cm}
\bibliographystyle{IEEEtran}

\bibliography{IEEEabrv,mybib}

\end{spacing}

\end{document}